\documentclass[manuscript, screen=true, natbib=false, nonacm, 11pt]{acmart}
\usepackage{hyperref}
\usepackage{cite}
\usepackage{amsmath,amsfonts}
\usepackage{algorithmic}
\usepackage{graphicx}
\usepackage{textcomp}
\usepackage{xcolor}
\def\BibTeX{{\rm B\kern-.05em{\sc i\kern-.025em b}\kern-.08em
    T\kern-.1667em\lower.7ex\hbox{E}\kern-.125emX}}

\newcommand{\eg}{\textit{e.g., }}
\newcommand{\ie}{\textit{i.e., }}

\usepackage{natbib}
\setcitestyle{acmauthoryear}
\begin{document}

\title{COOCK project Smart Port 2025 D3.1 \\``To Twin Or Not To Twin''}

\author{Randy Paredis}
\author{Hans Vangheluwe}
\author{Pamela Adelino Ramos Albertins}
\affiliation{
    \institution{University of Antwerp}
    \department{Department of Computer Science}
    \city{Antwerp}
    \country{Belgium}
}

\setcopyright{none}

\begin{abstract}
This document is a result of the COOCK project ``Smart Port 2025: improving and accelerating the operational efficiency of a harbour eco-system through the application of intelligent technologies''. 
It reports on the needs of companies for modelling and simulation and AI-based techniques, with twinning systems in particular. This document categorizes the purposes and Properties of Interest for the use of Digital Twins. It further illustrates some of the twinning usages, and touches on some of the potential architectural compositions for twins. This last topic will be further elaborated in a followup report.
\end{abstract}

\maketitle

\section{Introduction}
In 1970, the Apollo 13 mission was hit with disaster after an explosion in the oxygen tanks critically damaged their main engine. To make matters worse, the astronauts were about 330,000 km away from home, making it almost impossible to solve the issue from Earth. Luckily, NASA had a trick up their sleeve: the 15 training simulators were recalibrated to match the oxygen tank explosion. This allowed the engineers to come up with a solution for the issue and safely bring the astronauts home. To many researchers (including NASA), this feat of technology is considered the first Digital Twin (DT) \citep{apollo13,AllenNASA,boschert2016digital}.

\begin{figure}[htb]
    \centering
    \includegraphics[width=.85\textwidth]{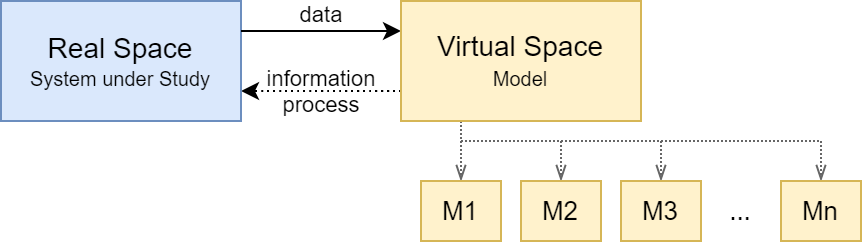}
    \caption{The ``conceptual model for PLM'', adapted from \citep{grieves2017digital}.}
    \label{fig:PLM}
\end{figure}

Yet, it wasn't until 2001, that the concept of ``Digital Twin'' was born. In a presentation with a specific focus on the aerospace industry, a conceptual model for Product Lifecycle Management (PLM) was shown (see figure~\ref{fig:PLM}), sprouting the idea behind DTs \citep{grieves2017digital}.
PLM focuses on following a company's products throughout their lifetime \citep{Stark2022}, which can be helped massively by using DTs.

While the DT concept can also be referred to as ``Virtual Digital Fleet Leader'', ``Information Mirroring Model'', ``Mirrored Spaces Model'' and ``Mirror Worlds'', the term ``Digital Twin'' in itself is the general consensus since 2010 \citep{singh2021digital}.
Since then, it has sprawled into many different domains and evolved to suit many specific needs of systems engineers. It is currently considered a cornerstone for Industry 4.0 (and by extension, Industry 5.0).

In the 2017 Gartner Research Hype Cycle for Emerging Technologies\footnote{\url{https://www.gartner.com/en/documents/3768572}}, the term ``Digital Twin'' appeared at the Innovation Trigger, yet in the 2018 edition of that same report\footnote{\url{https://www.gartner.com/en/documents/3885468}}, it had already moved to the Peak of Inflated Expectations. 
Surprisingly, later Gartner Hype Cycle reports
no longer mention DTs. DTs are omnipresent in industry (and a variety of other domains). This indicates that the concept has reached the Slope of Enlightenment, though possibly not yet the Plateau of Productivity, given that DTs and their variations (1) solve real problems and (2) are based on methods, techniques and architectures that were developed to maturity in the last decades.

Despite its maturity, there is currently no consensus for what a DT actually is.
\cite{Diakite2023} provides a theoretical description for a DT, based on class diagrams; and
\cite{Rumpe2021} counts 112 different definitions for DTs\footnote{\url{https://awortmann.github.io/research/digital_twin_definitions/}}, as a result of a systematic mapping study of 1471 unique publications \citep{DALIBOR2022111361}. While, in general, there is some consensus on these definitions, some are too domain-specific (like \cite{Steinmetz2018}), while others are incredibly vague (like \cite{yacob2019anomaly}). As a consequence, DT related terminologies like Digital Model, Digital Shadow, Digital Generator, Digital Avatar, Digital Thread, Digital Cockpit\ldots also suffer ambiguous definitions.

On top of that, one could argue that the DTs from the Apollo 13 are not actually \emph{digital}. Instead a highly complicated electromechanical device was used to \emph{physically} twin the spacecraft.
Even though some components may have been digital, the twin itself wasn't. We can refer to this kind of systems as ``Analog Twins'' (ATs).
The electromechanical device can also be referred to as a Cyber-Physical System (CPS).

\cite{Carreira2020} defines CPS as systems that consist of \emph{cyber} (as computerised implementations) and \emph{physical} components. The general idea is that the cyber and physical components influence each other in
such way that the cyber is able to cause the physical component to change state, and that the change, in turn,
will feed-back, resulting in a change of state on the cyber component.

We introduce the notion of ``\emph{twinning}'', bypassing the confusion of the many DT definitions, whilst at the same time also considering ATs.
When you have a representative, modelled counterpart of a System under Study (SuS), which is continually and continuously updated w.r.t. the state of the SuS, a ``twin'' appears. This twin should be calibrated and kept equivalent to the SuS w.r.t. a specific view and validity frame.
Some researchers also refer to the SuS as the ``Actual Object'' (AO). Similarly, the twin can be referred to as the ``Twin Object'' (TO).
In reality, the concept of ``twinning'' itself is not new.
In the antiquity, the ancient Greek (among others) used \textit{abax} (\eg sand tables) to ``twin'' the battlefields, allowing them to evaluate and plan troop movements. Nowadays, we still use sand tables (or equivalent systems) in military planning, psychology, education and a plethora of board games.
While not true for all use-cases, for some you can argue the sand tables are not just simulation models used for ``what if'' analysis, but are constantly kept ``in sync'' with the SuS, with the purpose of evaluating Properties of Interest (PoIs).

PoIs are a core concept for twinned systems. In \cite{Qamar2012}, a ``property'' is defined as the ``descriptor of an artifact''. It is a specific concern of an artifact (\ie a system), which is either logical (\eg the cruise controller in a ship is a PID-controller) or numerical (\eg there are 3 propellers to drive a specific vessel). Such properties can be \textit{computed} (\ie derived from other, related artifacts) or \textit{specified} (by a user). PoIs consider the specific system properties that are of concern to certain stakeholders. These can be used as formalizations of those stakeholders' goals.

The next sections of this report will mainly focus on a certain set of usages for twinning, as found in literature. They highlight the needs a user might have for creating a twin and how that might be encouraged.

\section{Categorization}
Given the overwhelming amount of literature on DTs, researchers have started to study and chart the purposes, methodologies and PoIs used in the engineering of DTs.
\cite{DALIBOR2022111361} gives a systematic mapping study, yielding a large feature model.
Similarly, in \cite{van2020taxonomy}, a taxonomy of DTs is constructed, based on 8 different dimensions.
\cite{JONES202036,kritzinger2018digital} mainly focus on the specific kinds of architectures they have encountered. In \cite{wanasinghe2020digital}, a set of business-specific classifications was made and \cite{Minerva2020,lim2020state} mainly discuss the kinds of technology and deployment.
Using \cite{DALIBOR2022111361}'s feature model as a starting point, also incorporating insights from other related research (including our own), a new, detailed set of aspects was constructed that relate to any number of twinning systems. Furthermore, this yields a fully detailed list of questions and topics that engineers should focus on when designing a twinned system in general.


\subsection{Purposes}\label{sec:goals}
First, let's discuss the goals and purposes w.r.t. (a set of) PoIs. These are generally goals that focus on the SuS. In essence, the SuS is the essential, central component that is to be analyzed.

\subsubsection{Design}\label{sec:design}
Many literature studies and taxonomies highlight that some twins are created \emph{before} the actual SuS. It is used to design all the required parts and already analyze it to ensure correct functionality. This is also called Variation Analysis \citep{Wang2018} or Virtual Prototyping \citep{poppe2019multi}. Design-Space Exploration allows engineers to explore any number of variations to check their accuracy w.r.t. some PoI.
While these techniques may help and enable twinning, in essence, they are normal Model-Based Systems Design/Engineering techniques that apply to any SuS (and not just twins).
For the purposes of this categorization, we will state that twinning can only appear as soon as \emph{both} the SuS and its model (\ie the twin) are present.

\subsubsection{Operation}
Most purposes for creating a twin have to do with the operation (\ie the at-runtime execution) of the overall system. We can consider a few sub-categories.

\paragraph{Data Allocation / Persistence}
In order for a twin to work, data needs to be allocated from the AO during its lifetime.
Some twins only use \emph{memorization} (\eg only keeping track of the current runtime information) \citep{SHARMA2022100383,santillan2019simulation}. Memorization should be present for any twin, otherwise it cannot be kept equivalent to the SuS.
The next step is to actually \emph{record data} for a longer period of time \citep{brockhoff2021process}. This maintains historical data and is able to make decisions based on what happened in the past.
A twin becomes intelligent as soon as we also maintain the context in which the data was obtained \citep{PADOVANO2018631}. This \emph{knowledge collection} enables inferencing and reasoning on past decisions in order to optimize future decisions.

\paragraph{Data Processing / Analysis}
The data that was obtained from the AO can be processed and analyzed in order to gain more insights into the system state (for both the AO and the TO).
The most simplistic analysis is \emph{validation}. Here, it is verified that the AO and the TO are accurate and represent the same kind of system \citep{Sharma2018RBFEABD,Grinshpun2016FromVT}. All models should be validly calibrated before their use, so in essence, validation should already be present in any twin.
Based on historical data, we might also do \emph{state estimation} in order to try and predict future situations \citep{GONZALEZ2020107371}.
We can take this to the next level by also doing \emph{behaviour/process prediction} \citep{brockhoff2021process} and/or \emph{forecasting} \citep{Gockel,Kosicka2017IntelligentSO}. Mainly using knowledge about the models and current state, we will predict a future for the system.
\emph{What-If Simulation} is also a common technique that is used to both \emph{explain} \citep{Rivera2022} and \emph{predict} \citep{flammini2021digital} situations in the system.

\paragraph{Observe / Monitor}
Some twins monitor the data in order to identify things that may have gone wrong.
The most famously used is \emph{anomaly/fault detection} \citep{yacob2019anomaly,Verriet2019}, where we just identify a drift between the AO and the TO.
Based on this drift, \emph{fault diagnosis} can be started to identify the reason thereof \citep{Verriet2019,Jain2020,Gonzalez2019}. Here, it can be detected that the AO's \emph{health} might have decreased \citep{GONZALEZ2019630}, or that some \emph{process parameters} are throwing a spanner in the works \citep{brockhoff2021process}.

\paragraph{Modify / Update}
According to \cite{kritzinger2018digital}, we can only call DTs ``Twins'' if there is a control signal from the TO to the AO. In essence, this is where the power of DTs comes from: having an external process update, \emph{optimize} or \emph{control} the AO.
The modification may be enabled using Design-Space Exploration and could be used to solve \emph{self-adaptation} \citep{bolender2021self,oreizy1999architecture,weyns2010forms}, \emph{self-control} \citep{GRAESSLER2018185}, \emph{self-healing} \citep{FENG2023119169}, \emph{self-reconfiguration} \citep{Mueller2021}, \emph{self-learning} \citep{moya2020physically} \ldots

\subsubsection{Visualization}
Most twins have a visual representation, such that a user is able to easily discern the current state of the system and/or make decisions based on this. While this visualization may be just that (\ie \emph{read-only}), it could also provide a set of controls for a user to make decisions (\ie \emph{read-and-write}).

\paragraph{Console / Dashboard}
A console is a text-based tool that either logs the current state, or allows you to browse this state \citep{lutters2019development}. A dashboard (also known as a Digital Cockpit \citep{brockhoff2021process}) is commonly a Graphical User Interface (GUI) that provides the same information, but in a more user-friendly manner. Dashboards may include plots, figures, tables and potentially a runtime debugger of the TO.

\paragraph{Animation} 
Many systems also show a graphical animation of the current state of the system. This may be in two dimensions (2D), using a plot \citep{feng2021incubator}, or a map \citep{paredis2021exploring}; or in three dimensions (3D) in a virtual world \citep{hopfner2021use}. 3D animations are commonly combined with Augmented Reality (AR), Virtual Reality (VR) or Mixed Reality (MR) approaches \citep{ElSaddik2018,Utzig2019}.

\subsubsection{Maintenance}
A twin \emph{evolves} over time. Similar to real-world CPSs, there may be wear and tear, changing requirements, system updates and changing technologies. Hence, it is important for a twin to be maintained throughout its lifetime.
Additionally, a TO may be able to deduce all these aspects of the AO and act upon them.

\paragraph{Predictive Maintenance}
Based on vendor specifications, past experiments and future predictions, it is possible to predict when a system will fail. \cite{Verriet2019} describes a smart lighting system for a building in which the twin can predict when the lights will break. When close to their end-of-life, it will automatically order a new light bulb, such that it can easily be maintained. Note that this example also shows that this maintenance process may not be fully automated.

\paragraph{Fatigue Testing / Damage Evaluation}
Instead of testing the sturdiness and/or wear and tear of a physical system, \emph{fatigue testing} \citep{gomez2020development} and/or \emph{damage evaluation} \citep{Utzig2019} translate this analysis (and its required preventive measures) to the virtual world.

\paragraph{Lifecycle Management}
As is the case for any system, the AO has an individual lifecycle that might have consequences for the state of the TO and all analyses that are happening \citep{Heithoff2023,Qin2022}.

\subsection{Quality Assurance}
Another aspect to study about twins is the quality they assure. How good the twinned system works. While most of this has to do with the deployment and implementation, focusing on this beforehand will result in a system that has a higher quality.

\subsubsection{Consistency}
We need to verify that the AO and the TO are kept in \emph{equivalence}. As per the original twinning description, this aspect needs to be present in every twin. We can monitor this using \emph{consistency monitoring} \citep{TALKHESTANI2018159} and by defining the allowed \emph{deviation}.

\paragraph{Synchronization}
A TO needs to be continually and continuously be kept in sync with the AO. While we can (in a perfect world) assume that the AO and the TO are not drifting, we can commonly not incorporate all aspects that cause drift in the model. In these situations, we need to re-sync whenever there is some deviation.
Alternatively, some technologies allow re-syncing every $x$ time, making sure the AO and TO are kept up-to-date \citep{MODONI2019472}. A downside of this approach is that this will require a lot of additional computations and that drift may not be detectable.

\subsubsection{Data Link}
Because a twinned system has some connection between the AO and the TO, the quality of the system will highly depend on this link itself. Additionally, the AO (and the TO) might have themselves some internal connections (\eg wiring, networks\ldots) that also should yield a high quality. Ideally, a high \emph{throughput} \citep{Sun2017}, high \emph{response time} \citep{Jain2020,Pargmann2018} and low \emph{latency} \citep{MASHALY2021299} are desired. This is also seen as ``\emph{data integration}'' \citep{URBINACORONADO201825,MASHALY2021299}.
Furthermore, any connection inside or between the AO and TO might need \emph{preprocessing} to reduce signal interference and sensor errors.

\subsubsection{Timing}\label{sec:timing}
The timing of a twinned system can be discussed in multiple dimensions. Either we can discuss the speed of execution of the TO (which can be \emph{slower-than-real-time}, \emph{real-time}, \emph{faster-than-real-time}, or \emph{as-fast-as-possible}) \citep{Diakite2023}. Depending on which timing the TO needs, the accuracy and precision of the twin will suffer. Some TOs cannot be executed in real-time (or faster), whereas others might not be able to slow down. In general, it is possible to state that the faster the updates can occur, the better the quality of the system becomes.
Alternatively, we can also discuss the timing of evolution of the overall system.

\subsubsection{Ilities}
\cite{Ilities} defines ``ilities'' as desired properties of systems, such as \emph{flexibility} or \emph{maintainability} (usually --but not always-- ending in “ility”), that often manifest themselves after a system has been put to use. These properties are not the primary functional requirements of a system's performance, but they typically involve wider system impacts. These ilities usually do not include factors that are always present in a system.
Important ilities that should be discussed (and analyzed) within the context of twinning are (but not limited to):

\paragraph{Efficiency}
How efficient is the overall system and the individual components?

\paragraph{Explainability} 
Can the system setup be used to explain the current state and its consequences?

\paragraph{Interoperability}
Describes how the twinned system can collaborate with other existing systems and how it can/may be used in a larger whole \citep{gross1999report}.
\cite{Traore2023} describes different kinds of interoperability and how to deal with them.

\paragraph{Fidelity}
There is no good consensus on what ``fidelity'' actually means. Most sources use it as a synonym for ``accuracy''. Yet, it is also used as ``level of polish and preparedness of a software product'' \citep{Refinery2016}, ``number of parameters transferred between the physical and virtual entities, their accuracy, and their level of abstraction'' \citep{JONES202036}, ``transmission performance'' \citep{shi2021new}, ``faithfulness'' \citep{oakes2023examining,rakove1996fidelity}, or a ``measure of realism'' \citep{gross1999report}.
For the purposes of this research, we will assume fidelity is equivalent to ``structural validity'' (\ie architectural correspondence / morphism between the model and the SuS) \citep{QUDRATULLAH20102216}.

\paragraph{Testability}
\cite{carnap1936testability} defines testablity as ``whether or not we know a method for testing''. Thus, we can ask ourselves if we can stress-test a system and/or if it can be used to run any number of scenarios. The main aspect here is to ensure there exists a method for testing the system within its production environment. Because the AO and the TO need to be valid and in sync, the TO can be used to test the AO beforehand \citep{Zhang2023}.

\paragraph{Availability}
What is the up-time of the twin?

\paragraph{Auditablity and Traceability}
Auditing might require a detailed description of what happened in the system. It might be required when errors occurred, or to check its safety (see later). For this, we need to maintain traces of every part of the execution and keep track of its traceability (which can be course-grained, or fine-grained \citep{Paredis2022FTGPM}).

\paragraph{Extensibility / Expansibility / Scalability / Elasticity / Dependability}
How many components can be added? Is the twinned system ``plug-and-play''?
Can we enlarge the overall system for a broader use-case? Will the system follow the current required demands? Do we upscale/downscale online (\ie at runtime), or offline?

\paragraph{Reliability / Stability}
Making sure that the system's execution can be trusted. For instance, having a large ``mean time to faillure'' (MTF) \citep{maxwell1978determination}.
Additionally, one can analyze how error-prone a system is. Techniques like \emph{Fault Injection} \citep{markwirth2021dynamic,Zhang2023} will help to verify this.

\paragraph{Maturity}
Which Technology Readiness Level (TRL) level can be assigned to the created system? \citep{DALIBOR2022111361}

\paragraph{Trust}
Is the person who built the twinned system to be trusted? Is it okay to add the system into a larger ecosystem of technologies? Can the constructed system be packaged and distributed to the outside-world? Can the TO trust the input communications from the AO \citep{ElSaddik2018}?

\subsubsection{Security / Safety / Robustness}
When building a system, sometimes there conditions that should be considered when building the TO, whilst not being ``theoretically'' required to be there. These compose (but are not limited to):

\paragraph{Federal Laws / Legal Safety}
When a government or other official instance prevents certain executions of a system. For instance, \cite{denil2012calibration} describes a power window that should stop closing when it gets a counter-force of 5N (to prevent it from chopping off limbs).

\paragraph{Physical Laws}
Some situations may be mathematically reachable (in a simulator), but ignore all laws of physics (\eg a robotic arm cannot rotate through itself).

\paragraph{Human Safety}
While some conditions may not be illegal, they should be prevented because of human safety. For instance, \cite{Paredis2021DT} describes that a human may not be in an industrial high-oven when it is turned on.

\paragraph{Fault Tolerance}
How much of the observed errors we will allow.

\paragraph{Privacy Enhancement}
Companies are commonly really interested in their Intellectual Property (IP), hence many techniques exist to ensure \emph{IP protection}. A common one is the usage of Functional Mock-Up Units (FMUs)\footnote{\url{https://fmi-standard.org/}}. Ideally, their system should be considered a ``black box'' that can be used and interacted with, without knowing the actual internal implementation(s).
It stands to reason that a company which builds a twin wants to keep the internal workings secret for the outside world. Furthermore, if this twin is used, you ideally don't want any hacker or malicious usage that could break a volatile system.

\subsection{Usage Contexts}
Another important aspect to study is when the twin is being used and in which context. \cite{kang2013variability} defines ``usage context'' as the set of circumstances where a system is operated in. It can also be seen as any contextual setting in which a product is deployed and used \citep{lee2010usage}.

\subsubsection{Sustainability}
The world we live in has a lot of battles left to fight. About 7 million people still live in poverty, preventing them the required access to education and employment. There is water scarcity and global warming ensures polluted air and a lot of greenhouse gasses.
With the popularization of Artificial Intelligence (AI) for the masses, even more pollution is being generated. According to \cite{FREITAG2021100340}, between 1.8 and 2.8\% of global greenhouse gas emissions is due to ICT and digital computations.
\cite{Michael2023} highlights that the biggest future for DTs appears when focusing on sustainability. Green IT is on the rise and twinning should jump on the bandwagon.
Some research is already being done by analyzing the energy efficiency of DTs \citep{bellis2022challenges} and sustainable manufacturing \citep{he2021digital}.

\paragraph{Footprint}
Additionally, we can consider the economical sustainability of having a twin. It might have a high maintenance cost, or high emissions. Even when it minimizes the footprint of the AO, the twin's cost should be minimized.

\subsubsection{Execution}
The execution of the TO can happen \emph{live} (\eg \emph{online}, given real-time access to sensor information and state data), or \emph{offline} (given past system execution traces). Notice that the choice of execution really links to timing discussed in section~\ref{sec:timing}. An online twin is what is commonly desired, but an offline one can help with debugging, calibration and validation of the model.

\subsubsection{User}
Who will be using the twin or the twinned system? Commonly, this is a \emph{manufacturer} when the twin is used to create and develop the AO \citep{Lutters2018PILOTPE,bolender2021self,rovzanec2021actionable,Mandolla2019}. When used at runtime, the user is a \emph{customer} \citep{Biesinger2018}. Alternatively, twinning ecosystems might ensure that a \emph{machine} is using the twin.

\subsubsection{Education}
Some twinned systems are used for \emph{education} \citep{um2017plug} or \emph{training} \citep{boschert2016digital}. They allow people to work with an actual system, whilst still active in a controlled environment. This makes teaching the usage of (for instance) heavy machinery a whole lot safer.

In essence, any twin can be used for education or learning about a complex system, even when they are not originally designed for it. Having enough data and system knowledge will enable understanding of a system. Adding a physical ``toy''-system\footnote{Which is not really a toy, but the actual system.} to that, \emph{kinaesthetic learners} (people who learn by doing \citep{Reese2007}) can even better understand the system.

\subsubsection{PLM Stage}
\begin{figure}[htb]
    \centering
    \includegraphics[width=0.95\textwidth]{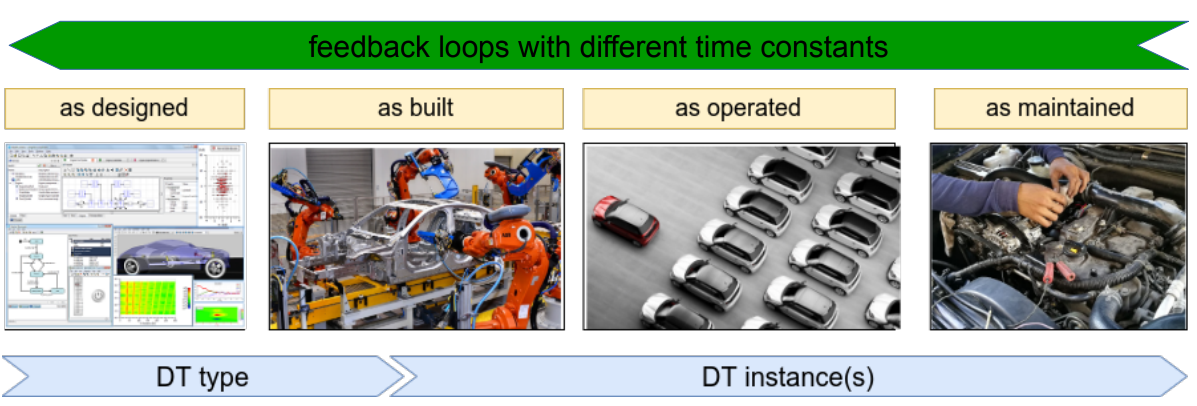}
    \caption{Digital Twin PLM stages}
    \label{fig:DT-stages}
\end{figure}

The origins of DTs can be traced back to PLM \citep{grieves2017digital}. It would hence make sense to maintain the equivalence between PLM and DTs (or twinning in general). Many PLM systems are used to allow twinning communication too \cite{DALIBOR2022111361}. Note that a twin is not required to be part of a single stage, but it might span multiple. Figure~\ref{fig:DT-stages} shows the DT stages graphically.

\paragraph{As Designed}
Section~\ref{sec:design} describes the \emph{design} stage of PLM. The AO does not exist yet, so technically we cannot yet consider twins.
\cite{grieves2017digital} calls this the ``\emph{DT type}'', given that it is not a twin yet, but rather an idea for a twin.
Here, models can be used to construct and analyze a prototype of the product.

\paragraph{As Built}
The \emph{building} (or \emph{manufacturing}) stage focuses on how the system is constructed. It uses a (virtual) prototype in order to start constructing the actual SuS. We can start considering the prototype as a twin.
Twinning is a common practice that appears whilst manufacturing an AO \citep{Lutters2018PILOTPE}.
Additionally, there exist a plethora of twins for the manufacturing process itself \citep{bolender2021self,rovzanec2021actionable,Mandolla2019}.


\paragraph{As Operated}
Most twins are used in the \emph{operation} or \emph{usage} stage. Here, the AO is running in parallel to the TO. Maybe the AO starts to deviate (because of its specific usage context), allowing the TO also to deviate.

\paragraph{As Maintained}
Sometimes the product needs repairs, or must be \emph{maintained} in one way or another. Twinning is commonly used in this stage too \citep{lehner2021aml4dt,DALIBOR2022111361}.


\subsubsection{Reuse}
Some twins are created for a very specific use-case, yet others might be able to be implemented in many different scenarios and contexts. Reuse focuses on how a twin might reappear in a different scenario. \cite{hong2021reproducibility} and \cite{repeatabilityETC} define \emph{repeatability} (the same team can produce the same results with the exact same experimental setup), \emph{replicability} (a different team can produce the same results with the exact same experimental setup) and \emph{reproducibility} (a different team can produce the same results with a different experimental setup) in the context of scientific experimentation.

When a twin is used for optimization of an existing system, allowing the AO to perform better, \emph{system reconditioning}\footnote{\url{https://sbt-alliance.com/what-is-retro-commissioning/}} might occur.
Finally, there is also \emph{design reuse}, in which the current system design is reused to tackle a different problem \citep{Landahl2018TowardsAD}. Via a well chosen system architecture, design reuse can be implicitly enabled.

\section{Architectures}
\begin{figure}[htb]
    \centering
    \includegraphics[width=.85\textwidth]{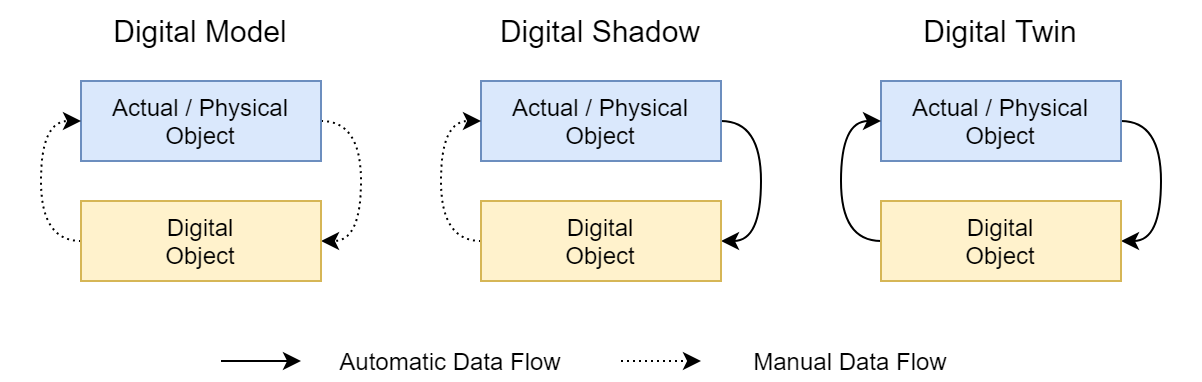}
    \caption{Digital Twin subcategories according to \cite{kritzinger2018digital}.}
    \label{fig:Kritzinger}
\end{figure}

The most basic DT architecture is described in \cite{grieves2017digital}. Based on the level of data integration between a physical object (\ie a SuS, or ``asset'') and its digital counterpart (\ie a virtual model), \cite{kritzinger2018digital} proposes three subcategories: \textit{Digital Model}, \textit{Digital Shadow} en \textit{Digital Twin}. This is illustrated in figure~\ref{fig:Kritzinger}. When there is no automated information flow between both entities, the system is a \textit{Digital Model}. When the flow from the physical object to the virtual model is automated, it is called a \textit{Digital Shadow}. The \textit{Digital Twin} only appears when all connections (within the range of operation) are automated.
\cite{tekinerdogan2020systems} also introduces a \textit{Digital Generator}, which is the dual of the \textit{Digital Shadow}. The latter paper also describes a meta-model and a set of (software-inspired) architectural patterns that can be used in the development of DTs.

\begin{figure}[htb]
    \centering
    \includegraphics[width=0.6\textwidth]{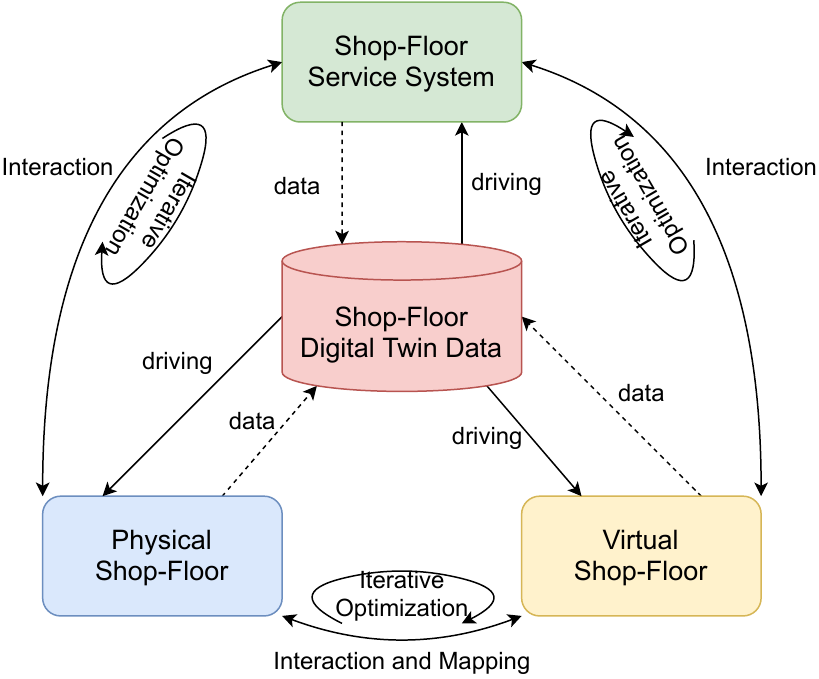}
    \caption{5D Architecture model for DT on the Shop-Floor, adapted from \cite{tao2017digital}.}
    \label{fig:5D}
\end{figure}

Earlier work, \cite{Paredis2021DT}, extended the architecture of \cite{kritzinger2018digital} by stating that at least a third component would be required in the actual realization of DTs. This component partially handles the communication and objective observation of the other components.
\cite{tao2017digital} goes further and describes five main components in a DT architecture of a shop-floor. This is referred to as the 5D architecture model \citep{van2021models}. This 5-Dimensional (5D) model is illustrated in Figure~\ref{fig:5D}.
In the context of production systems, \cite{tao2017digital} describes five main components in a DT architecture of a shop-floor.
Similar to most architectures, there is a (physical) SuS (\ie ``Physical Shop-Floor''), which is linked to a (digital) model (\ie ``Virtual Shop-Floor''). The ``Shop-Floor Service System''  contains all the external applications (\eg dashboards, anomaly detection, fault tolerance services, \ldots) that require information from one (or both) of the other components. In order to facilitate the communication between the SuS, the model and the services, the ``Shop-Floor Digital Twin Data'' is used. This component drives the SuS and the virtual model. Finally, the fifth dimension is presented in the connections between all components.

\subsection{Twinning System Components}
All these structural designs, as well as \cite{JONES202036} allows us to describe the individual components and aspects related to the \emph{structure} of a twinned system. Note that each concept described in this section might have multiple interacting instances in the overall twinned system.

\subsubsection{Actual Object}
The Actual Object (AO) or SuS is the system for which we would like to have a twin. It should solve a problem on its own, while its twin aims for one of the goals (see section~\ref{sec:goals}).
This AO is also referred to as ``Real Space'' \citep{grieves2017digital}, ``Physical Shop-Floor'' \citep{tao2017digital}, ``Physical Counterpart'' \citep{DALIBOR2022111361}, ``Physical Object'' \citep{kritzinger2018digital,becker2021conceptual,paredis2021exploring} or ``Physical Entity'' \citep{JONES202036,negrin2021using}, ``Physical Twin'' \citep{david2023digital}, ``Twin of Interest'' \citep{Diakite2023}, or ``Actual System'' \citep{tao2022digital}.

Note that ``Physical'' often recurs here. This is based on the idea that the SuS should always be a purely physical system. This is not always the case, given that twins of the following systems also exist:

\paragraph{Biochemical Entity}
Considers people (as a complex biochemical system), animals, organs, plants, medicine\ldots\ 
\cite{david2023digital} discusses Cyber-Biophysical systems within the context of DTs. Other sources focus on greenhouses \citep{howard2021greenhouse}, agriculture \citep{tekinerdogan2020systems,verdouw2017digital}, chemical reactions \citep{silber2023accelerating}, or life sciences \citep{lifescienceDT,ElSaddik2018}.

\paragraph{Social Entity}
Considers people and animals within a social context. It mainly focuses on the interaction instead of the individual. Examples could be employees in a business structure \citep{GRAESSLER2018185}, citizens in a city \citep{Traore2023}, or actors/animals in the case of motion capture\footnote{\url{https://en.wikipedia.org/wiki/Motion_capture}}.

\paragraph{Electromechanical Entity}
Concerns cars \citep{Barosan2020DevelopmentCenter}, machines \citep{Mandolla2019}, robots \citep{paredis2021exploring,marah2023MADTwin,Soccer_robots}\ldots
Within the context of Industry 4.0, this kind of entity is the main application for twinning.

\paragraph{Solid Entity}
An arbitrary physical object \citep{ElSaddik2018,DALIBOR2022111361}. For instance, \cite{bonney2021digital} discusses how a twin of a metal shelf can be constructed.

\paragraph{System of Systems}
A system in which there are multiple sub-systems that may also have twins themselves. They enable twinning ecosystems in any dimension. It also allows us to construct Higher-Order Twinning Hierarchies (HOTH) via smart combining of these twins.
Complete factories \citep{Biesinger2018} are a good example of such systems.

\paragraph{Workflows / Business Processes}
Concerns all workflow-based systems (\ie a system in which a sequence of activities is studied). Commonly, DTs are build for business processes \citep{RambowHoeschele2018CreationOA}.

\paragraph{Software}
Anything in the virtual world. There could exist twins for running software \citep{Heithoff2023}, running simulations, virtual worlds (\ie BIM, CAD\ldots) \citep{ANGJELIU2020106282} and data (\ie historical traces, experiment results\ldots).

\subsubsection{Twin Object}
Conversely, we will not be focusing on purely Digital Twins, but instead also consider the so-called Analog Twins.

\paragraph{Biochemical Entity}
Considers people (as a complex biochemical system), animals, organs, plants, medicine\ldots, just like before. For instance, \cite{topuzoglu2019finding} describes how slime molds were used to construct a map for the city of Izmir.
Given that it is generally close to impossible to influence a biochemical entity, allowing it to be not-so-equivalent to the AO. For this reason, very little twins exist with this specification.

\paragraph{Social Entity}
There is also the usage of people and animals within a social context, in which the focus lies on the interaction aspect. The human is not considered biologically. The most typical example here is acting. Actors that play historical figures, medical simulants, or family composition activities in psychology\ldots can all \emph{technically} be considered ``twins''.

\paragraph{Electromechanical Entity}
Concerns analog computers that exist mainly in the physical world. They might contain digital aspects, but they are not typically seen as ``digital'' on their own.
The Apollo 13 Space Capsule Training Simulator \citep{apollo13} is also this kind of entity.

\paragraph{Solid Entity}
A physical object that is commonly influenced by a human, or stands on its own. The oldest example of this is a Sand Table, which have been used since the antiquity and are still being used (albeit mainly in movies/entertainment) for military planning and wargaming.
Antwerp Zoo teaches children about deforestation and its impact on the tamarin population using a miniature model of the rainforest and decommissioned fire-hoses that represent roads or highways.
Architects build scale models of houses to test their structural soundness, capacity, light influence, \ldots

\paragraph{Digital Entity}
Appears whenever the \emph{full} twin is in the virtual space. This can be a simulation \citep{moya2020physically}, software, AI \citep{FENG2023119169}, or pure data \citep{Utzig2019}. All DT applications have this entity as model (if not, they are wrongly categorized).

\subsubsection{Experiment Manager}
\cite{wohlin2012experimentation} states that experiments  are  launched  when  we  want  control  over  the  situation  and  want to  manipulate  behavior  directly,  precisely  and  systematically.
Thus, an experiment can be seen as an execution of a twin such that one or more PoIs can be analyzed. The experiment managers set up, identify and manage how the data attained from the AO and TO can be used to satisfy a specific goal (\ie through a PoI) for which the twin was created.

\subsubsection{Services}
\cite{tao2017digital}'s ``\emph{Shop-Floor Service System}'' reappears in a lot of reference architectures as the set of \emph{services}. It is the set of external (potentially interacting) processes (either virtual, or physical) that uses data from the AO and the TO in order to achieve their goal. 
Commonly, a twin's PoIs, goals and purposes define the kinds of services that can be executed. Examples are dashboards, anomaly detection, fault injection, historical data services, observing, planning/strategizing\ldots

\subsection{Twinning Architecture}
Besides focusing on the individual components of a twin, we could also focus on the kind of architecture that the overal system will yield.

\subsubsection{Product Type}
The product type defines how the AO and TO work in terms of multiplicity. In essence: how many twins will the overall system contain?

\paragraph{Single System}
The most basic product type appears when only considering a single twin. In this case, there is a simple architectural structure and we will commonly also focus on a specific use-case (given the undoubtedly limited scope of problems that that twin could solve).

\paragraph{Hetereogeneous Group of Twins}
Also considered an ``\emph{ecosystem}'' of twins in which multiple TO instances have an immediate interaction with each other. This overall system is required to be orchestrated, similar to cosimulation \citep{BRECHER2021833}.
There are multiple reasons for having multiple twins. Each of the twins might have different \emph{PoIs}, \emph{goals} or \emph{views} on the same AS. There might be different \emph{levels of abstraction/detail}, or we might focus on the different \emph{PLM stages} (\eg a design model might not be needed anymore during runtime). Additionally, different \emph{formalisms} might be used, as per the Multi-Paradigm Modelling (MPM) methodology \citep{Mosterman2004}. For redundancy reasons, one might also use different \emph{copies} of the same system.

\paragraph{Hetereogeneous Group of Systems}
Conversely, the AO might in itself consist of multiple systems that work together to reach a common goal. In some situations, this group of AOs can be considered a multi-agent system \citep{KARADUMAN2023106478}.
In essence, such a system can be simplified to having a single AO by combining all sub-systems (and their interactions) into a single AO. There might be twins for each of the sub-systems, which will also be executed in parallel, ensuring validity on the instance level, as well on the overall level.

\paragraph{Hierarchical Composition}
Because a twin can be created for a system, but a system might also contain a twin, we can perfectly hierarchically decompose twins. Think of a factory, where every machine has a twin, where the factory itself is given a twin and that system might also get a twin focusing on the business processes. We consider these systems Higher-Order Twinning Hierarchies (HOTH).
Each level of the HOTH might contain a combination of any of the previously stated architectures. For instance, an AT linked to another AT (AT-AT), bridging the gap between two twinned systems, may be part of a HOTH when we build a DT to mimic the AT-AT's behaviour.

\subsection{Historical Information Service}
The Experiment Managers communicate with a special service, the \textit{Historical Information Service} (HIS), which stores all of the historical information (data, models, experiments, configuration, measurements, \ldots). The HIS also enables access to this data, which may be useful for certain PoIs that focus/rely on historical information. In essence, the HIS acts like a \textit{bloackboard system} for storing past data.

\section{Conclusions}
The twinning paradigm appears everywhere, in a plethora of domains and applied to any number of research questions. We can analyze their usages and their architectures in order to solve large-scale problems.
However, a DT is not a "silver bullet". It will not solve all problems the world faces, because we cannot simply account for all scenarios the real-world accounts for \citep{ibrion2019risk}.
This document describes the usages and contexts of twinning and which problems they can answer, mainly focusing on their PoIs. It highlights the user needs for twinning. But it is pertinent to know that a twin can only be used correctly once it was given a validity frame.

There exist many different usages and applications of DTs in literature and this document has tried to map them. When building a twin, many questions should pop up for the system engineers to keep in mind whilst constructing their twins.

\begin{acks}
This research was partially supported by Flanders Make, the strategic research center for the manufacturing industry. Additionally, we would like to thank the Port of Antwerp-Bruges for their explanations of internal functionalities of the nautical chain.
\end{acks}

\bibliography{main.bib}

\end{document}